\documentclass[a4paper, 14pt, conference,onecolumn]{ieeeconf}      

\IEEEoverridecommandlockouts                              

\usepackage{graphics} 
\usepackage{epsfig} 

\usepackage{eurosym}
\usepackage[latin1]{inputenc}
\usepackage[frenchb]{babel}
\newtheorem{remarque}{Remarque}[section]

\usepackage{amsmath} 
\usepackage{amssymb}  
\usepackage{tabularx}
\usepackage{amsmath}
\usepackage{balance}

\newcounter{subassumption}[asu]

\makeatletter
\renewcommand{\p@subassumption}{\theasu}
\makeatother

\usepackage{rotating,booktabs,multirow}
\makeatletter
\let\NAT@parse\undefined
\makeatother
\usepackage{hyperref}
\usepackage{graphicx}

\usepackage[latin1]{inputenc}
\usepackage{color}
\usepackage{epsfig}
\usepackage{graphicx}
\usepackage[hang]{caption2}
\usepackage[normalsize,it,hang]{subfigure}
\DeclareGraphicsExtensions{.jpg,.png,.jpg,.jpg}

\title{\LARGE \bf
Fouille de données et segmentation de chroniques par extrema : considérations préliminaires \\ { \normalsize Data mining and time series segmentation via extrema: preliminary investigations}}

\author{Michel Fliess$^{1, 2}$, C\'{e}dric Join$^{2, 3}$  
\thanks{$^{1}$LIX (CNRS, UMR 7161), \'Ecole polytechnique, 91128 Palaiseau, France. {\tt \small Michel.Fliess@polytechnique.edu } } 
\thanks{$^{2}$AL.I.E.N. (ALg\`ebre pour Identification \& Estimation Num\'eriques), 7 rue Maurice Barr\`{e}s, 54330 V\'{e}zelise, France. \newline
        {\tt \small \{cedric.join, michel.fliess\}@alien-sas.com}}
\thanks{$^3$CRAN (CNRS, UMR 7039)), Universit\'{e} de Lorraine, BP 239, 54506 Vand{\oe}uvre-l\`{e}s-Nancy, France. \newline
        {\tt\small Cedric.Join@univ-lorraine.fr}}
 }

\begin{document}
\maketitle
\thispagestyle{empty}
\pagestyle{empty}

{\bf  R\'{E}SUM\'{E}: }
\rm
{\em La segmentation des chroniques est l'un des outils de fouilles des donnés. On l'aborde, ici, en prenant des extrema locaux pour points d'importance perceptuelle (PIP). Un théorème de décomposition additive des chroniques, d\^{u} à Cartier et Perrin, et des techniques algébriques d'estimation des dérivées, déjà utiles en automatique et signal, permettent d'évacuer le brouillage d\^{u} aux fluctuations rapides de toute chronique. Des illustrations numériques valident notre approche et soulignent l'importance du choix du seuil de détection.} \\

{\bf  ABSTRACT$:$ }
{\em Time series segmentation is one of the many data mining tools. We take here local extrema as perceptually interesting points (PIPs). The blurring of those PIPs by the quick fluctuations around any time series is treated via an additive decomposition theorem, due to Cartier and Perrin, and algebraic estimation techniques, which are already useful in automatic control and signal processing. Our approach is validated by several computer illustrations. They underline the importance of  the choice of a threshold for the extrema detection.} 
\\~\\

{\bf MOTS-CL\'{E}S:}
\rm{\em chroniques, segmentation, point d'importance perceptuelle (PIP), estimation algébrique.}\\

{\bf KEYWORDS$:$}
\rm{\em time series, segmentation, perceptually important points (PIPs), algebraic estimation.}\\~\\

\section{Introduction}

\begin{figure*}[!h]
\centering%
{\epsfig{figure=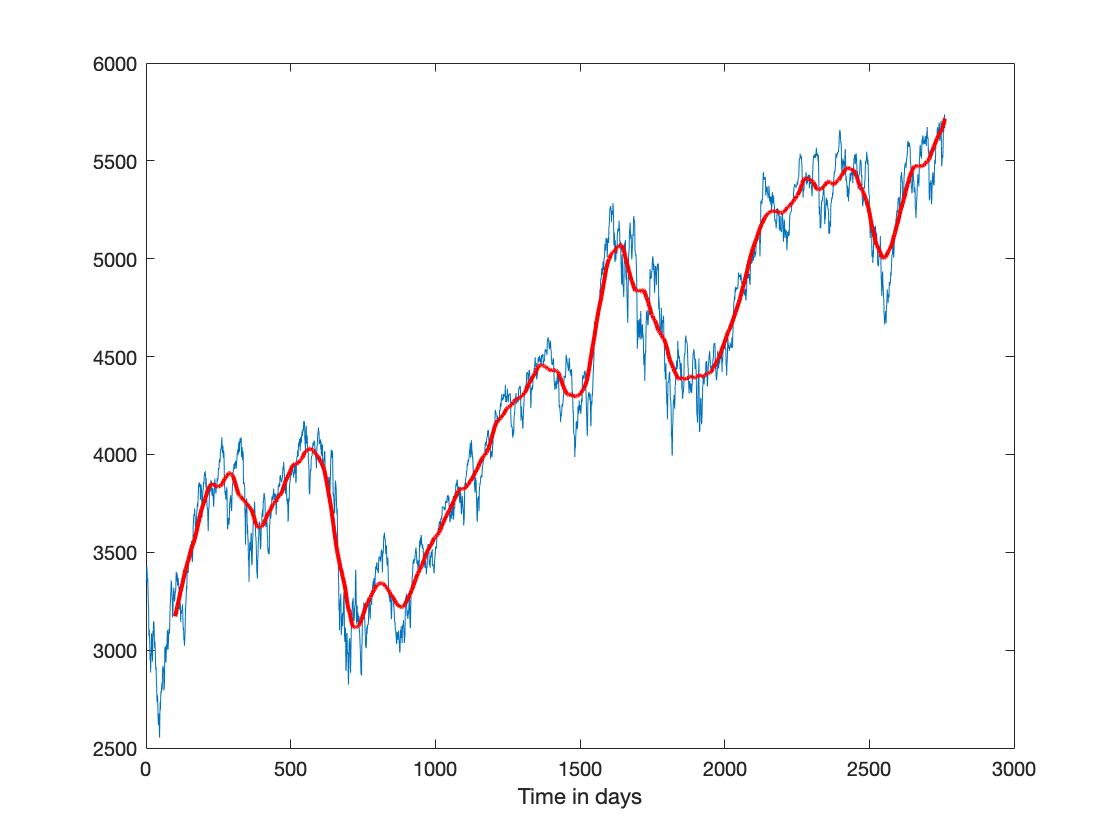,width=1.0\textwidth}}
\caption{CAC 40. En bleu: valeur quotidienne du CAC 40 du 1er janvier 2009 au 31 décembre 2019. En rouge: moyenne (glissante).}\label{Si}
\end{figure*}

Les chroniques, ou \og time series \fg, sont omniprésentes dans la fouille de données, ou \og data mining \fg. Parmi les multiples méthodes utilisées, on privilégie ici la \emph{segmentation} (voir, par exemple, \cite{abo}, \cite{cho}, \cite{duran}, \cite{duran1}, \cite{duran2}, \cite{esling}, \cite{keogh}, \cite{kim}, \cite{lee}, \cite{novel}, \cite{lu}, \cite{marti}), c'est-à-dire une approximation par un nombre fini de segments de droite. De nombreuses études concrètes l'ont déjà utilisée dans des domaines variés: hydrométéorologie (\cite{hubert}), économie (\cite{cheong}, finance (\cite{wan,wan1}), mouvement animal (\cite{animal}, vision par ordinateur (\cite{liu}), efficacité énergétique des machines outils (\cite{seevers}), collecte de données par drones aquatiques  (\cite{castellini}), $\dots$  Prendre pour bouts des segments des \emph{points d'importance perceptuelle} (\emph{PIP}), ou \og  \emph{perceptually important points}\fg, consolide ce point de vue  (voir, par exemple, \cite{fu,fu2}), qui l'est encore davantage si ces PIP sont des extrema locaux (\cite{pratt,fink1,fink2,yin}).  La détection des PIP est brouillée par les fluctuations rapides de toute chronique. Comment isoler des extrema locaux sur la ligne bleue de la figure \ref{Si}? Les publications existantes ignorent, semble-t-il, cette difficulté majeure. Cet article présente un cadre mathématique et algorithmique afin de détecter ces extrema locaux en dépit des oscillations. Quelques exemples académiques sur la réduction de dimension, préalable à toute fouille de données, illustrent notre propos.


Notre approche des chroniques, qui s'appuie sur (\cite{solar}), consacré à la prévision et à la gestion du risque pour l'énergie photovolta\"{\i}que, est née en ingénierie financière (\cite{trend,hatt}). Elle a été utilisée pour la prévision du trafic autoroutier (\cite{abouaissa}) et des ressources nécessaires en informatique nuagique, ou \og cloud computing \fg \ (\cite{cloud}). Le pivot en est un théorème dû à \cite{cartier}: toute chronique $X$ satisfaisant une hypothèse très lâche, s'écrit $X = E(x) + X_{\rm{fluctuat}}$, où $E(x)$ est la \emph{moyenne}, ou \emph{tendance} (\og \emph{trend} \fg), et $X_{\rm{fluctuat}}$ les fluctuations rapides autour de $0$. La ligne rouge de la figure \ref{Si} correspond à la moyenne.  Le théorème de Cartier et Perrin s'exprime dans le langage de l'\emph{analyse non standard}, introduit par  \cite{robinson} et outil merveilleux pour l'intuition (voir, par exemple, \cite{lobry}). On ne recherche pas les extrema sur $X$, mais sur $E(X)$, plus lisse. La détection d'un extremum est liée à l'annulation de la dérivée d'ordre $1$. On y parvient grâce aux techniques d'\emph{estimation algébrique} (\cite{easy,mboup}) pour traiter du cas bruité, connu comme important et difficile (\cite{lanczos}). Si la dérivée est proche de $0$ durant un certain temps, la chronique est remplacée par un plateau, c'est-à-dire un segment horizontal. Des techniques algébriques analogues ont déjà été employées pour détecter des ruptures, ou \og change-points \fg  (\cite{change}). 

Cette communication est organisée comme suit. Les rappels du § \ref{rappel} sont consacrés aux chroniques et à l'estimation algébrique. On présente au § \ref{illustration} des illustrations numériques, qui soulignent l'importance du seuil de détection des extrema. Quelques prolongements sont proposés au § \ref{conclusion}.

\section{Rappels sur les chroniques et l'estimation algébrique} \label{rappel} 

\subsection{Chroniques et analyse non standard}\label{series}
L'\emph{analyse non standard}, qui repose sur la logique mathématique, fournit un sens précis aux notions d'infiniment petit et d'infiniment grand. On recommande la présentation par \cite{nelson1} (voir aussi \cite{nelson2}, \cite{diener1,diener2}), conceptuellement plus simple.

Soit l'intervalle $[0, 1]$, avec échantillonnage infinitésimal, comme souvent en analyse non standard,
\begin{equation*}\label{sampling}
\mathfrak{T} = \{ 0 = t_0 < t_1 < \dots < t_\nu = 1 \}
\end{equation*}
où $t_{i+1} - t_{i}$, $0 \leq i < \nu$, est {\em infinitésimal}, c'est-à-dire très petit. 
\begin{remarque} \label{rem}
Les notions d'infiniment petit ou grand sont des idéalisations mathématiques. On doit considérer, en pratique, un laps de temps d'une seconde (resp. heure) comme très petit par rapport à une heure (resp. mois). C'est pourquoi l'analyse non standard s'applique au monde réel.
\end{remarque}

Une chronique $X$ est une fonction $\mathfrak{T} \rightarrow \mathbb{R}$. 
La {\em mesure de Lebesgue} sur ${\mathfrak{T}}$ est la fonction
$\ell$ définie sur ${{\mathfrak{T}}} \backslash \{1\}$ par
$\ell(t_{i}) = t_{i+1} - t_{i}$. La mesure de tout intervalle $[c, d]
\subset \mathfrak{T}$, $c \leq d$, est sa longueur $d -c$.  L'\emph{intégrale} sur $[c, d]$ de la chronique $X(t)$ s'écrit
$$\int_{[c, d]} Xd\tau = \sum_{t \in [c, d]} X(t)\ell(t)$$
$X$ est $S$-{\em integrable} si, et seulement si, l'intégrale $\int_{[c, d]} |X| d\tau$ est
\emph{limitée}, c'est-à-dire non infiniment grande, et, si $d - c$
is infinitésimal, $\int_{[c, d]} |X| d\tau$ l'est aussi.
$X$ est $S$-{\em continue} en $t_\iota \in {\mathfrak{T}}$ si, et seulement si, $f(t_\iota) \simeq f(\tau)$ quand $t_\iota \simeq
\tau$ (on écrit $a \simeq b$ si $a - b$ est infinitésimal).
$X$ est {\em presque continue} si, et seulement si, elle est
$S$-continue sur ${\mathfrak{T}} \setminus R$, où l'ensemble $R$ est {\em
rare}\footnote{$R$ est \emph{rare}
(\cite{cartier}) si, pour tout réel standard $\alpha > 0$, il existe un ensemble interne $A \supset R$ tel que $\ell(A) \leq \alpha$.}.
$X$ est \emph{Lebesgue intégrable} si, et seulement si, elle est
$S$-intégrable et presque continue.

La chronique $X$ est {\em rapidement fluctuante}, ou {\em oscillante}, autour de $0$, si, et seulement si, 
elle est 
\begin{itemize}
\item $S$-intégrable,
\item $\int_A {\mathcal{X}} d\tau$
est infinitésimal pour ensemble {\em quadrable}\footnote{Un ensemble est
\emph{quadrable} (\cite{cartier}) si sa frontière est rare.}.
\end{itemize}

\textbf{Théorème de Cartier-Perrin}: Si la chronique $X$ est $S$-intégrable, il y a décomposition additive\footnote{Voir \cite{lobry} pour une présentation plus abordable} (\cite{cartier}):
\begin{equation}\label{decomposition}
\boxed{X(t) = E(X)(t) + X_{\tiny{\rm fluctuat}}(t)}
\end{equation}
où
\begin{itemize}
\item la \emph{moyenne} $E(X)(t)$ est Lebesgue-intégrable,
\item $X_{\tiny{\rm fluctuat}}(t)$ est rapidement fluctuante.
\end{itemize}
La décomposition \eqref{decomposition} est unique à une quantité additive infinitésimale près.
\begin{remarque}
Cette moyenne, ou \emph{tendance}, souvent désignée par le mot anglais \og \emph{trend} \fg, est à rapprocher du lissage par \emph{moyenne glissante} ou \og \emph{moving average} \fg \ (voir, par exemple, \cite{melard}). On la retrouve aussi en \emph{analyse technique} (voir, par exemple, \cite{bechu,technical,tsinas}), branche de l'ingénierie financière. Très différent est le sens habituel de \emph{trend} dans la littérature sur les chroniques (voir, par exemple, \cite{enders}).
\end{remarque}

\subsection{Estimation algébrique des dérivées}\label{estimation}
Le § \ref{series} démontre que l'on peut passer au continu si les données sont en nombre suffisant et la période d'échantillonnage petite. La dérivée est, en fait, estimée sur la moyenne et non sur les données brutes.



Soit la fonction polyn\^{o}miale de degré $1$
$$p_1 (t) = a_0 + a_1 t, \quad t \geq 0, \quad a_0, a_1 \in \mathbb{R}$$ 
Le calcul opérationnel classique (voir, par exemple, \cite{yosida}) permet de réécrire $p_1$ 
$$P_1 = \frac{a_0}{s} + \frac{a_1}{s^2}$$ 
D'où en multipliant par  $s^2$
\begin{equation}\label{1}
s^2 P_1 = a_0 s + a_1
\end{equation}
et en dérivant par rapport à $s$, qui, dans le domaine temporel,
correspond à la multiplication par $- t$:
\begin{equation}\label{2}
s^2 \frac{d P_1}{ds} + 2s P_1 = a_0
\end{equation}
Le système triangulaire (\ref{1})-(\ref{2}) d'équations fournit $a_0$, $a_1$. On se débarrasse de $s P_1$, $s^2 P_1$, an $s^2 \frac{dP_1}{ds}$, c'est-à-dire, dans le domaine temporel, des dérivées par rapport au temps, en multipliant les deux membres des équations (\ref{1}) et (\ref{2}) par $s^{ - n}$, $n \geq 2$. Les intégrales itérées correspondantes, qui, en pratique, sont remplacées par des filtres linéaires numériques, sont des filtres passe-bas: elles atténuent les oscillations parasites. Une fenêtre temporelle courte suffit pour estimer avec précision $a_0$, $a_1$. 

La généralisation aux polyn\^{o}mes de degré supérieur est évidente, donc aussi celle aux développements de Taylor tronqués. Renvoyons à \cite{mboup} pour plus de détails.

\begin{remarque}
Les manipulations algébriques précédentes sont employées avec succès en ingénierie (voir, par exemple, \cite{sira}). Le passage en $0$ de la dérivée première a été utilisée, en particulier, en traitement du signal (\cite{fedele}).
\end{remarque}

\section{Illustrations}\label{illustration}

\subsection{Polyn\^{o}mes par morceaux}
\begin{figure*}[!h]
\centering%
\subfigure[\footnotesize En bleu: signal. En rouge: segmentation ]
{\epsfig{figure=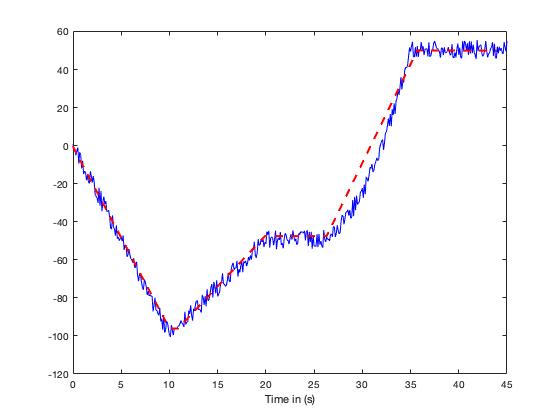,width=.7\textwidth}}
\subfigure[\footnotesize Véritable dérivée (noir - -), dérivée estimée (bleu), seuils (rouge - -), passage en $0$ (noir *)]
{\epsfig{figure=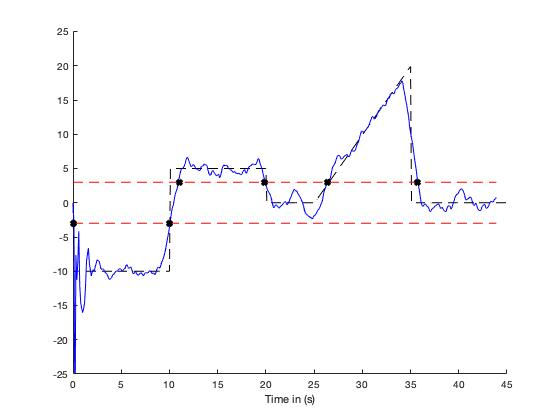,width=.7\textwidth}}
\caption{Polyn\^{o}mes par morceaux}\label{S1}
\end{figure*}

Soit $y(t)$ défini par
$$
\begin{cases}
y(t)=-10t+0.5 0 \quad 0\leq t< 10\\
y(t)=5t-99.5 \quad 10\leq t< 20\\
y(t)=-49.5 \quad 20\leq t< 25\\
y(t)=t^2-49.5 \quad 25 \leq t< 35\\
y(t)=50.5 \quad 35\leq t \leq 45\\
\end{cases}
$$
altéré par un bruit uniformément réparti entre $-5$ et $5$.
Soit $0.1$s la période d'échantillonnage. La 
segmentation obtenue est présentée figure \ref{S1}-(a). L'estimation de dérivée l'est, selon le § \ref{estimation}, avec un polynôme de degré $2$ sur une fenêtre de $40$ échantillons. La figure \ref{S1}-(b) en atteste la qualité. On note la détection des plateaux. La segmentation est peu représentative durant la branche parabolique, de courte durée heureusement\footnote{Voir, à ce sujet, le point 1 du § \ref{conclusion}.}.


\subsection{Sinuso\"{\i}des}
\begin{figure*}[!h]
\centering%
\subfigure[\footnotesize En bleu: signal. En rouge: segmentation ]
{\epsfig{figure=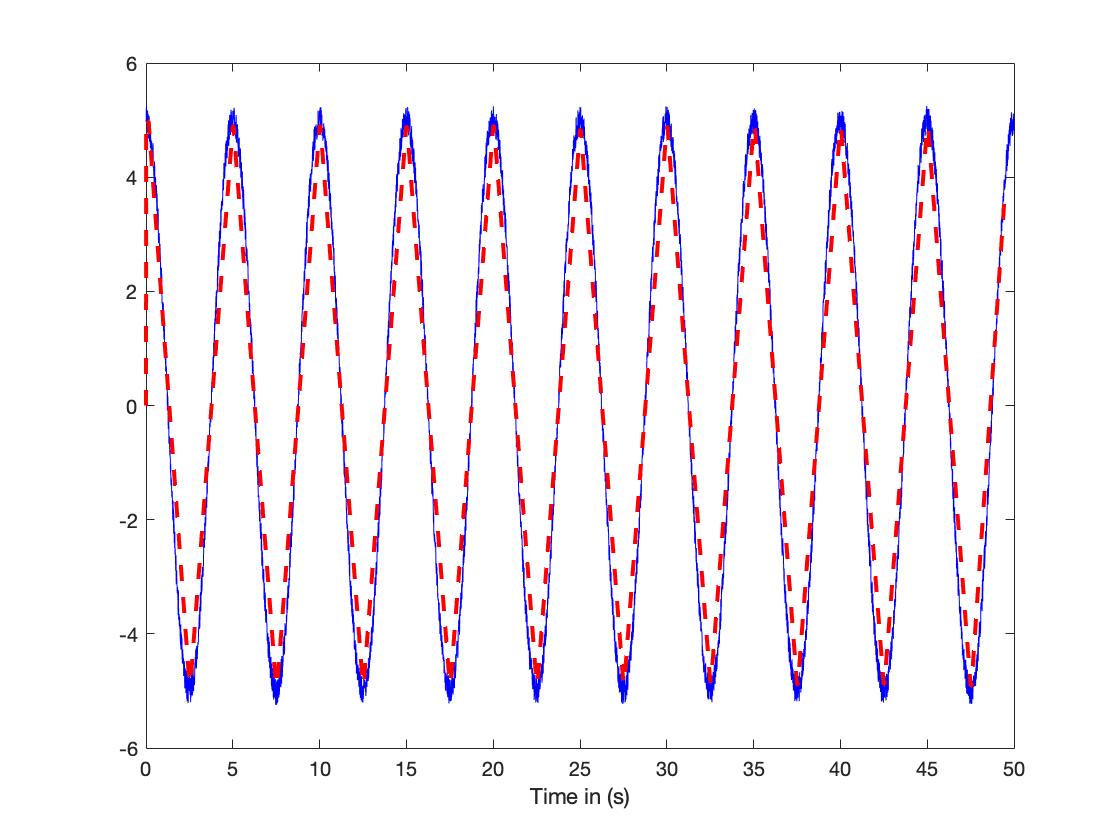,width=.492\textwidth}}
\subfigure[\footnotesize Véritable dérivée (noir - -), dérivée estimée (bleu), seuil (rouge - -), passage en $0$ (noir *)]
{\epsfig{figure=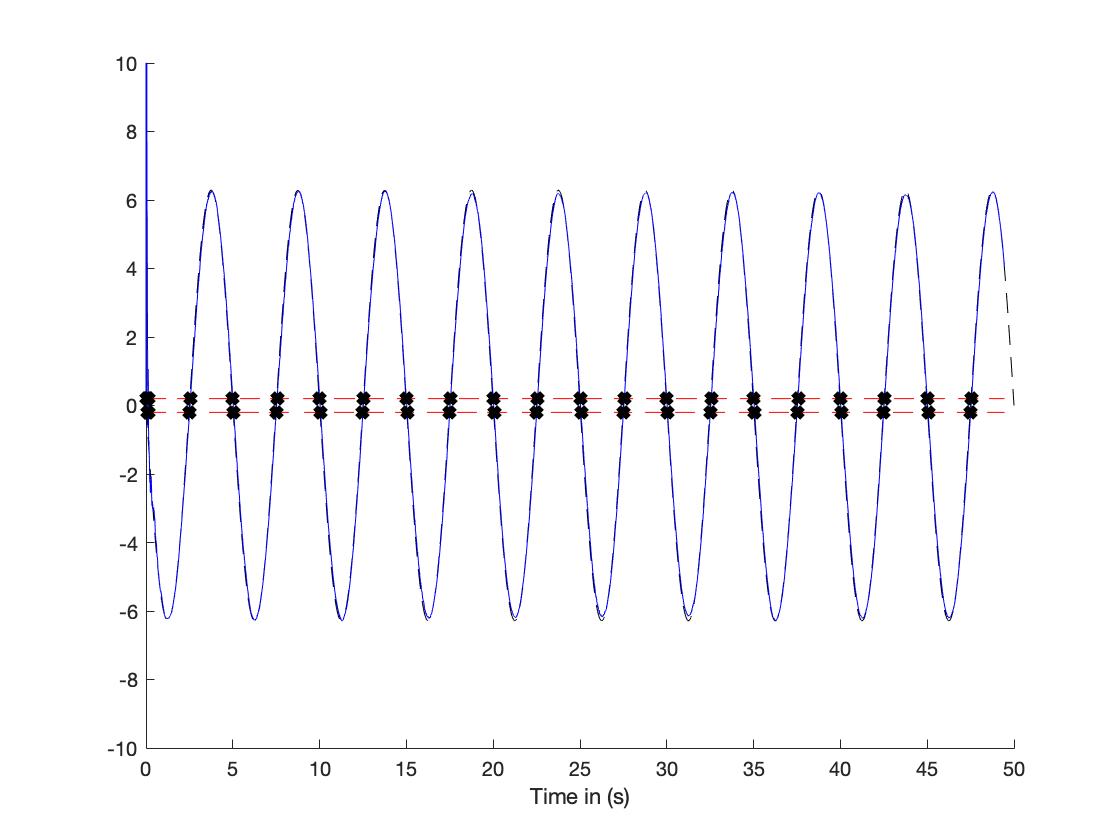,width=.492\textwidth}}
\subfigure[\footnotesize Zoom de (a) ]
{\epsfig{figure=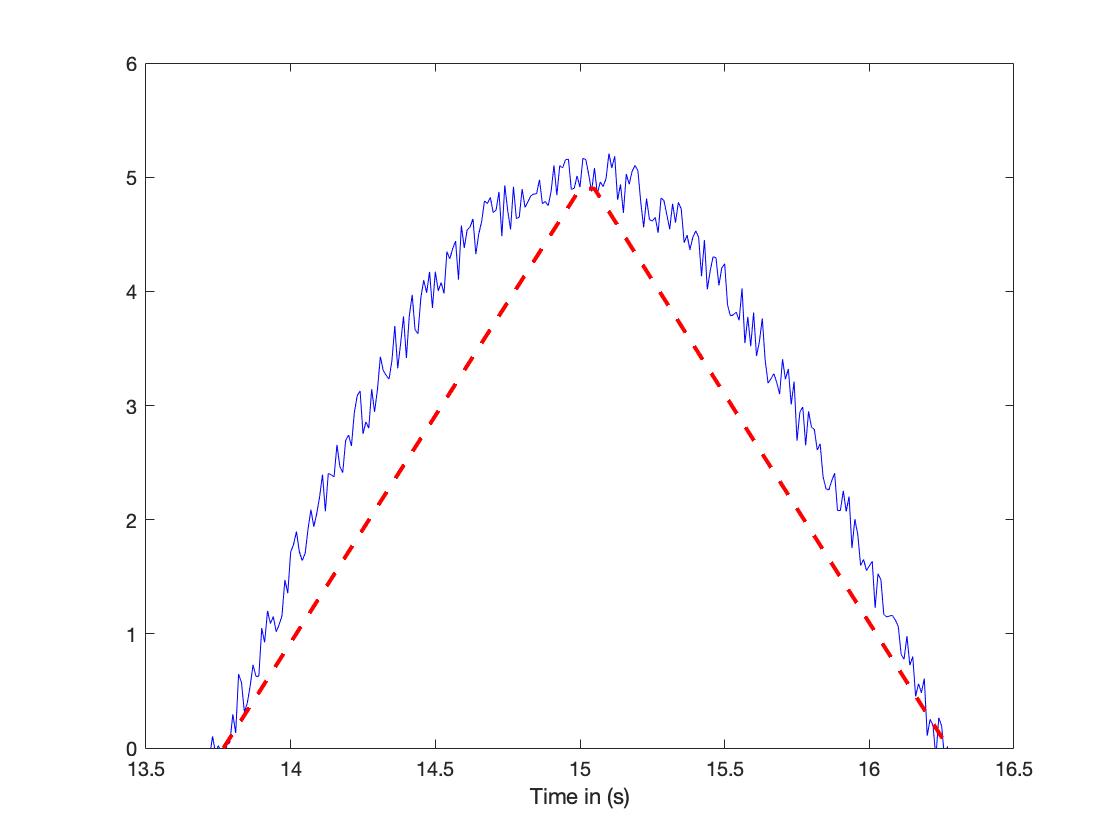,width=.492\textwidth}}
\subfigure[\footnotesize Zoom de (a) ]
{\epsfig{figure=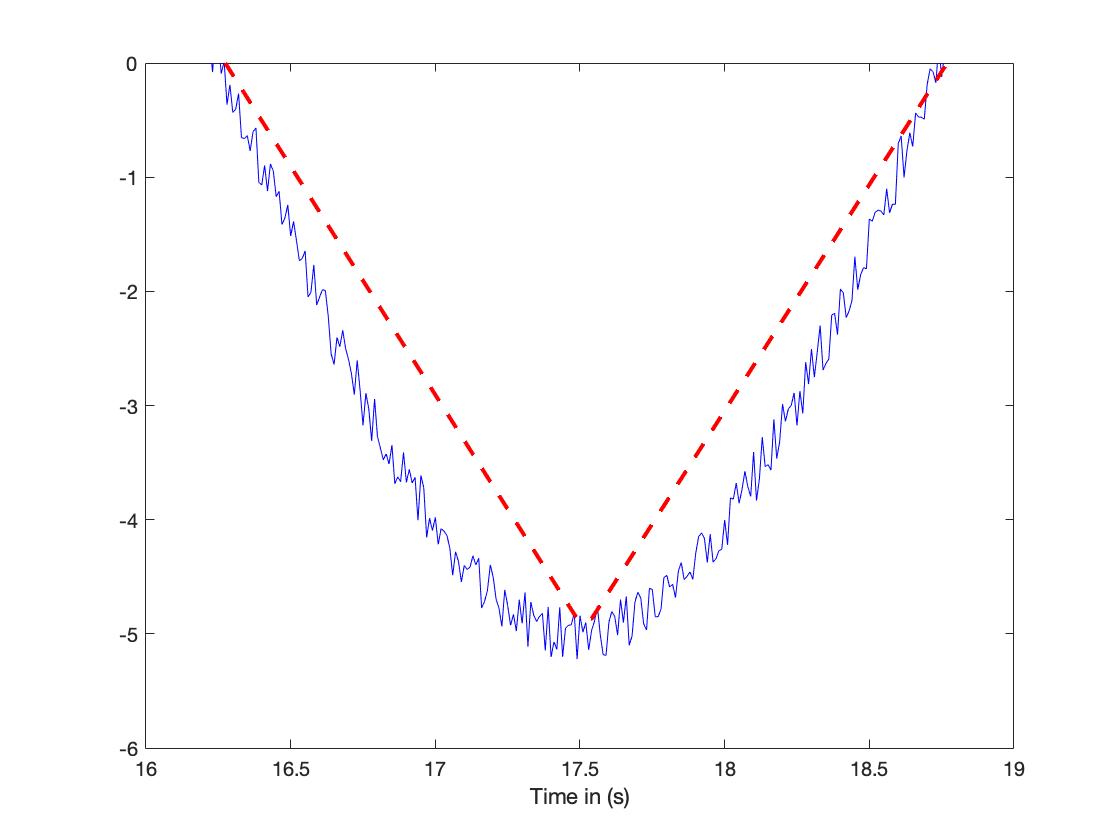,width=.492\textwidth}}
\caption{Sinusoïde: fréquence élevée}\label{S2}
\end{figure*}

\begin{figure*}[!h]
\centering%
\subfigure[\footnotesize  En bleu: signal. En rouge: segmentation ]
{\epsfig{figure=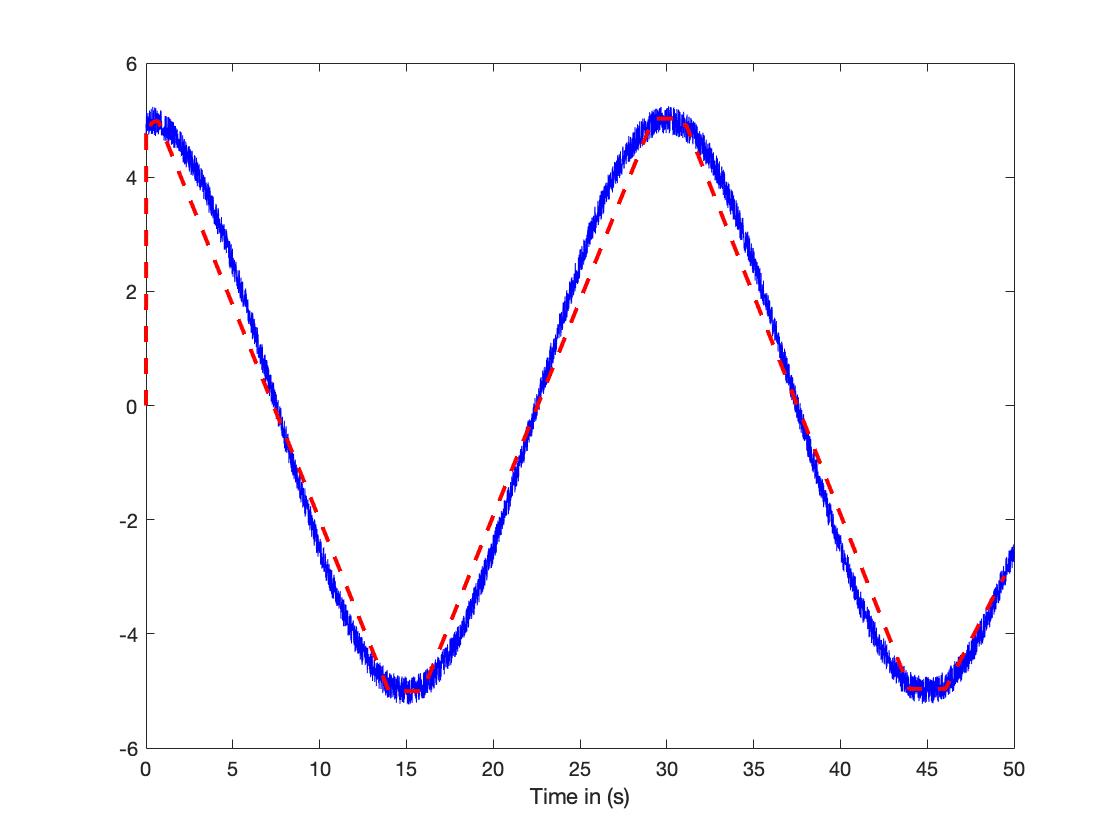,width=.492\textwidth}}
\subfigure[\footnotesize Véritable dérivée (noir - -), dérivée estimée (bleu), seuil (rouge - -), passage en $0$ (noir *)]
{\epsfig{figure=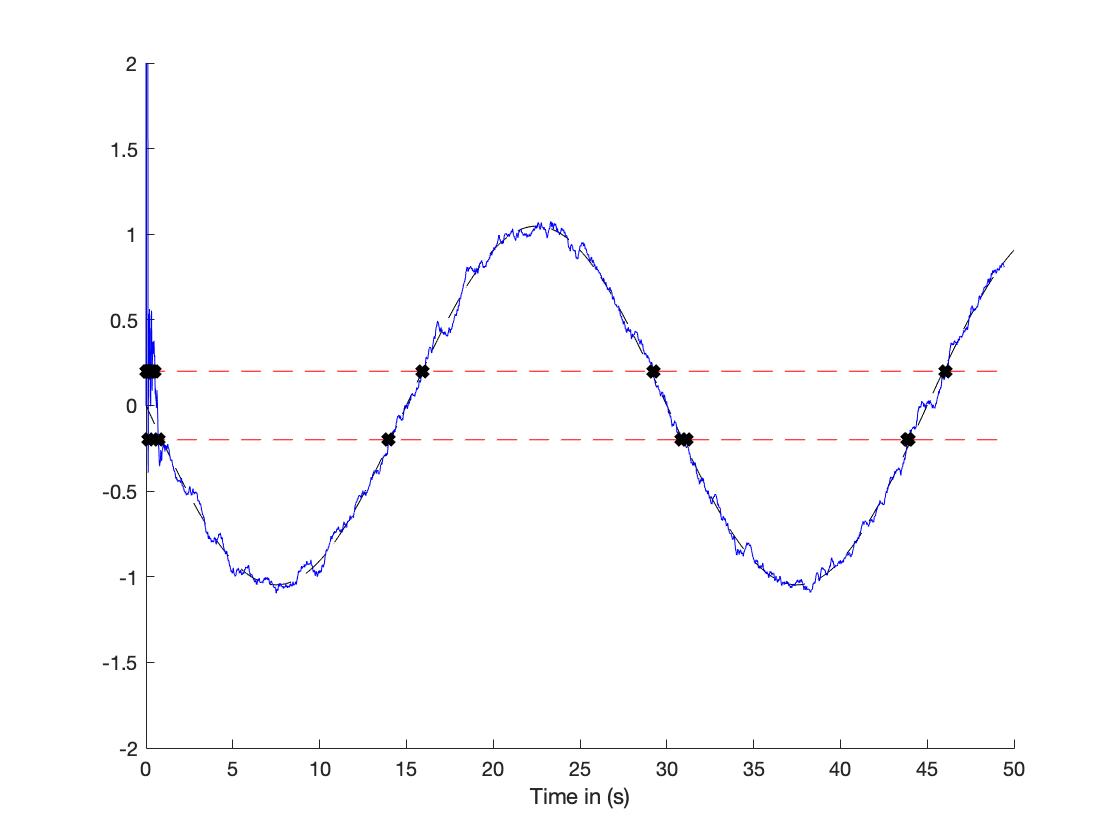,width=.492\textwidth}}
\subfigure[\footnotesize Zoom de (a) ]
{\epsfig{figure=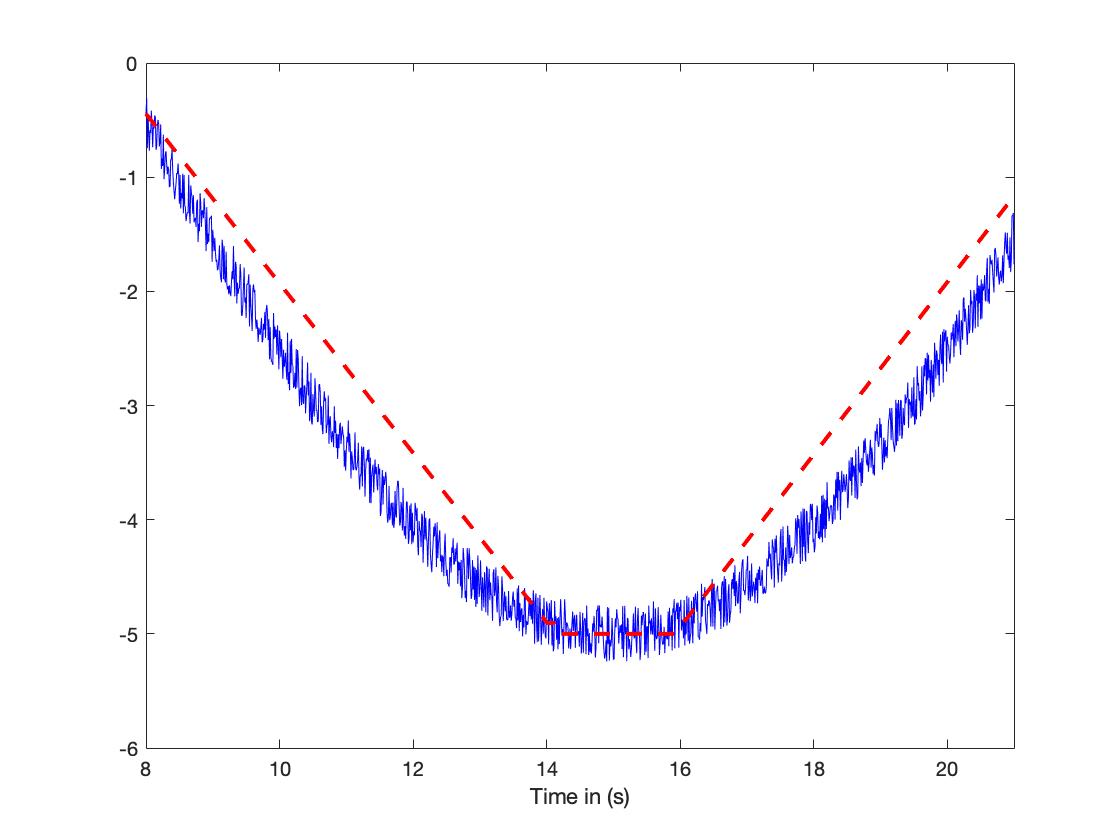,width=.492\textwidth}}
\subfigure[\footnotesize Zoom de (a) ]
{\epsfig{figure=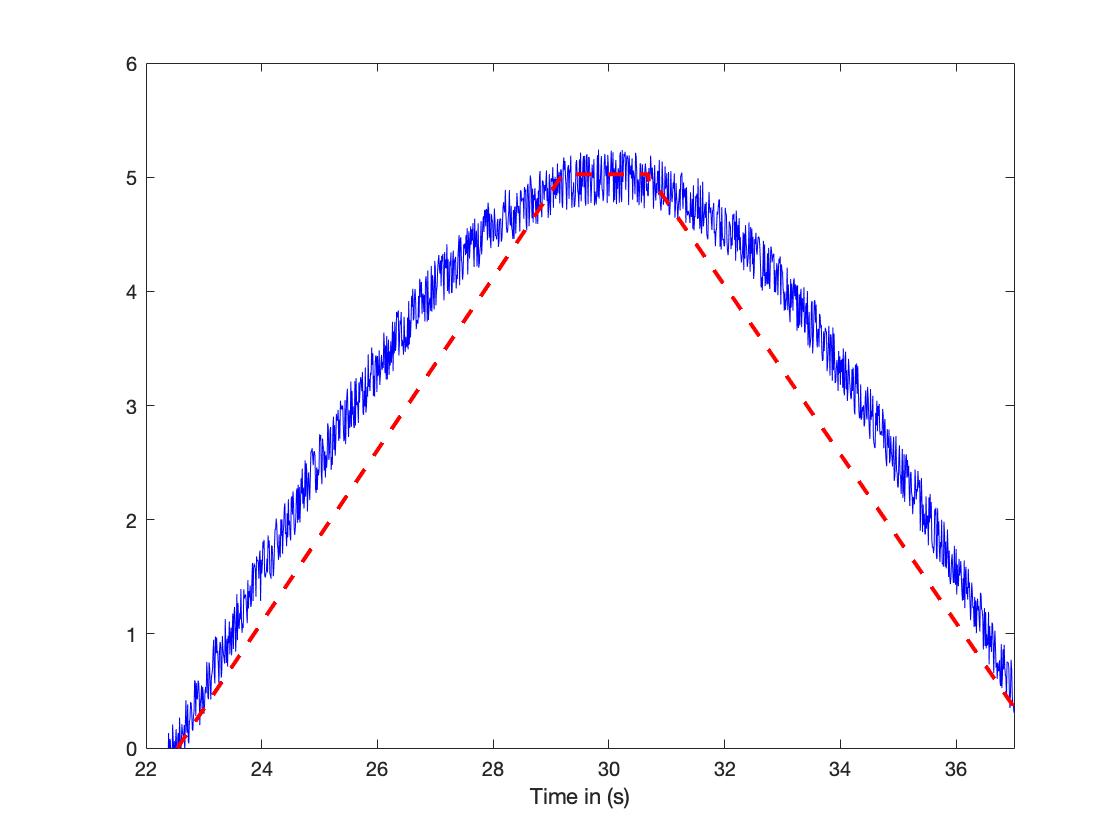,width=.492\textwidth}}
\caption{Sinusoïde: fréquence basse}\label{S3}
\end{figure*}

Soient deux sinuso\"{\i}des de fréquences $5$s et $30$s, échantillonnées toutes les $0.01$s (voir figures  \ref{S2}-(a) et  \ref{S3}-(a)). Elles sont altérées par un bruit uniformément réparti entre $-0.25$ et $0.25$.
La dérivée est estimée avec un polynôme de degré $2$, sur une fenêtre glissante de durée $2$s.
On détecte les extrema avec un seuil égal à $0.2$ (voir figures \ref{S2}-(b) et \ref{S3}-(b)). Avec la petite fréquence, ce seuillage conduit à un plateau (figures \ref{S3}-(c) \& \ref{S3}-(d)), mais pas avec la grande (figures \ref{S2}-(c) \& \ref{S2}-(d)).



\subsection{CAC 40}
\begin{figure*}[!h]
\centering%
\subfigure[\footnotesize En bleu: chronique. En rouge: segmentation ]
{\epsfig{figure=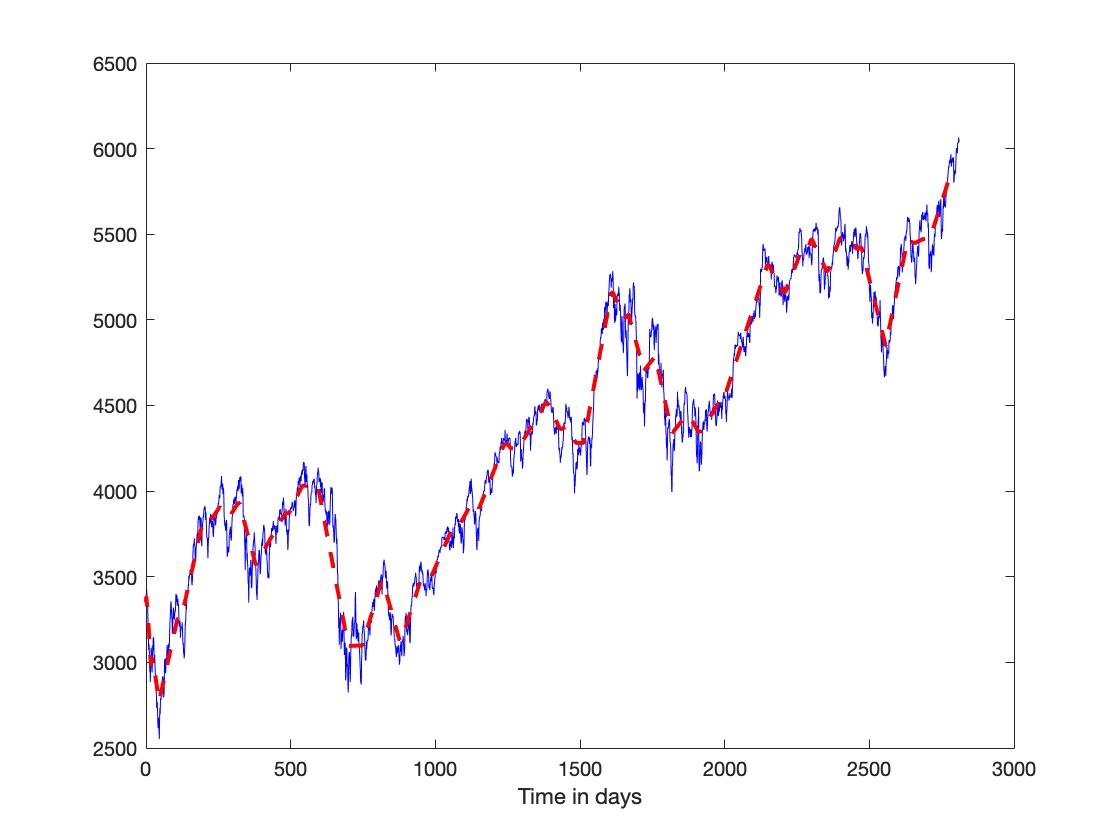,width=.492\textwidth}}
\subfigure[\footnotesize Dérivée estimée (bleu), seuil (rouge - -), passage en $0$ (noir *)]
{\epsfig{figure=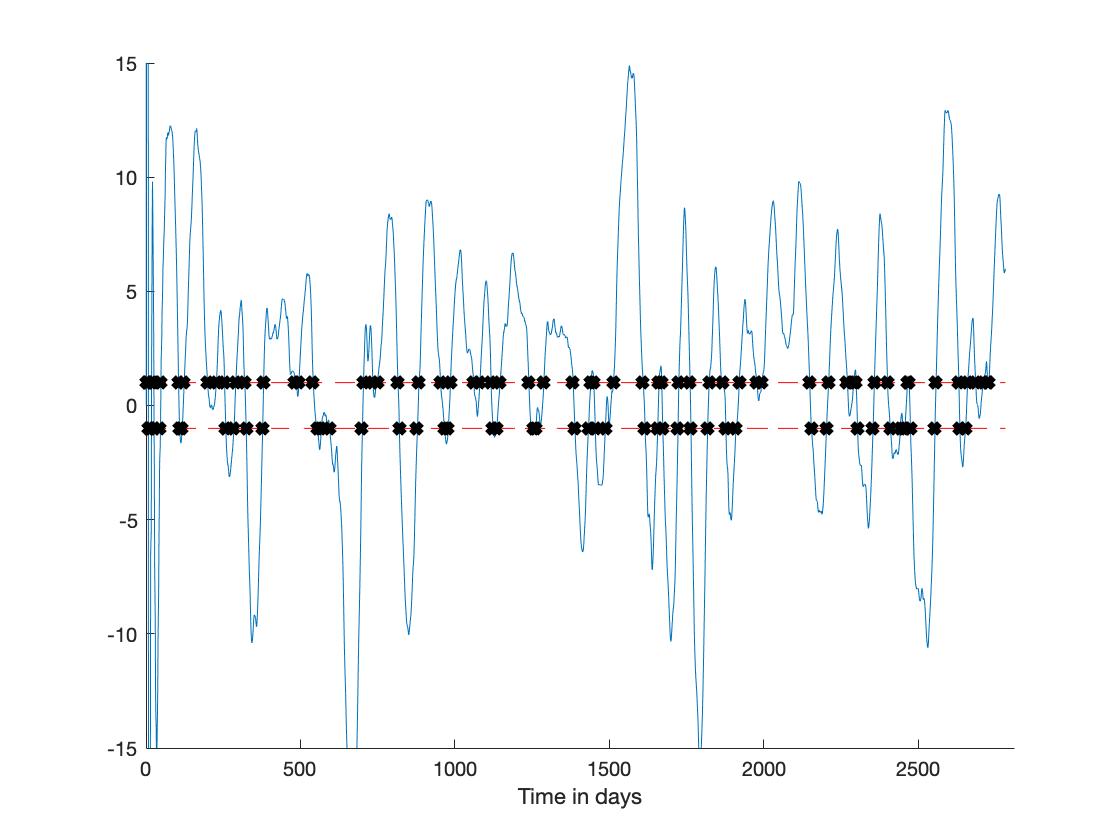,width=.492\textwidth}}
\subfigure[\footnotesize Zoom de (a) ]
{\epsfig{figure=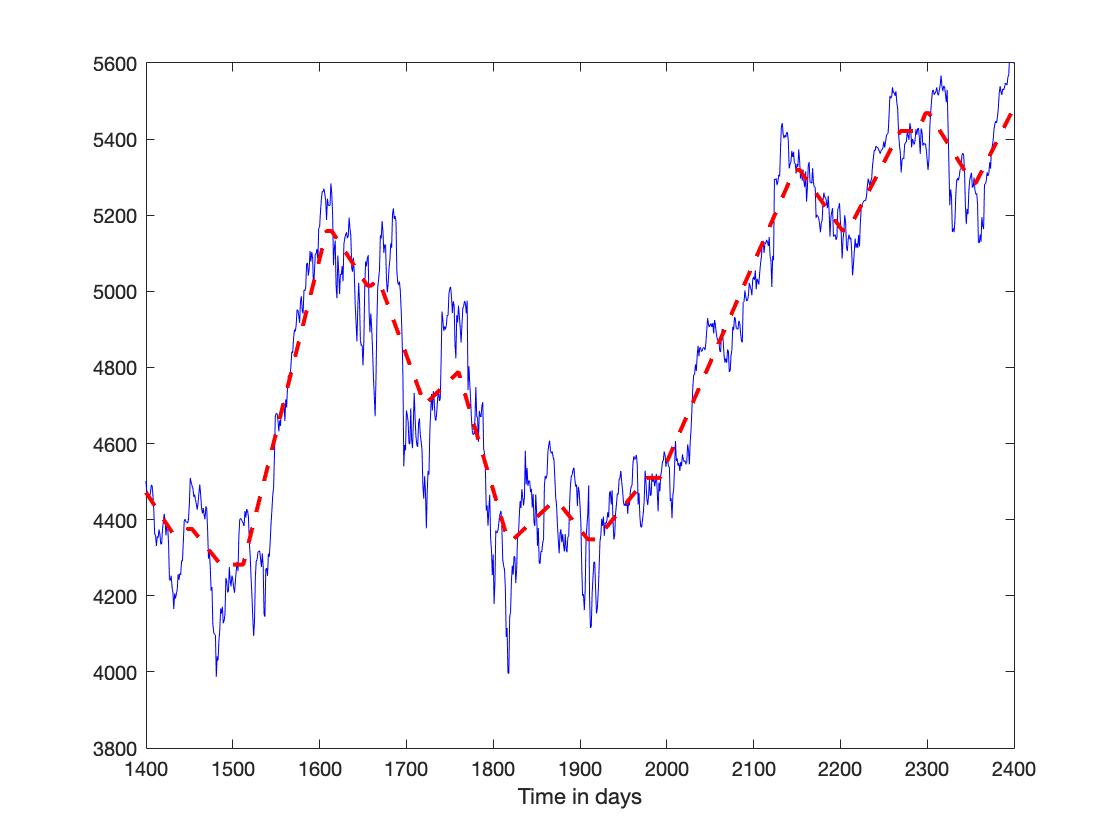,width= .95\textwidth}}
\caption{CAC 40: Seuil bas}\label{S41}
\end{figure*}

\begin{figure*}[!h]
\centering%
\subfigure[\footnotesize En bleu: chronique. En rouge: segmentation ]
{\epsfig{figure=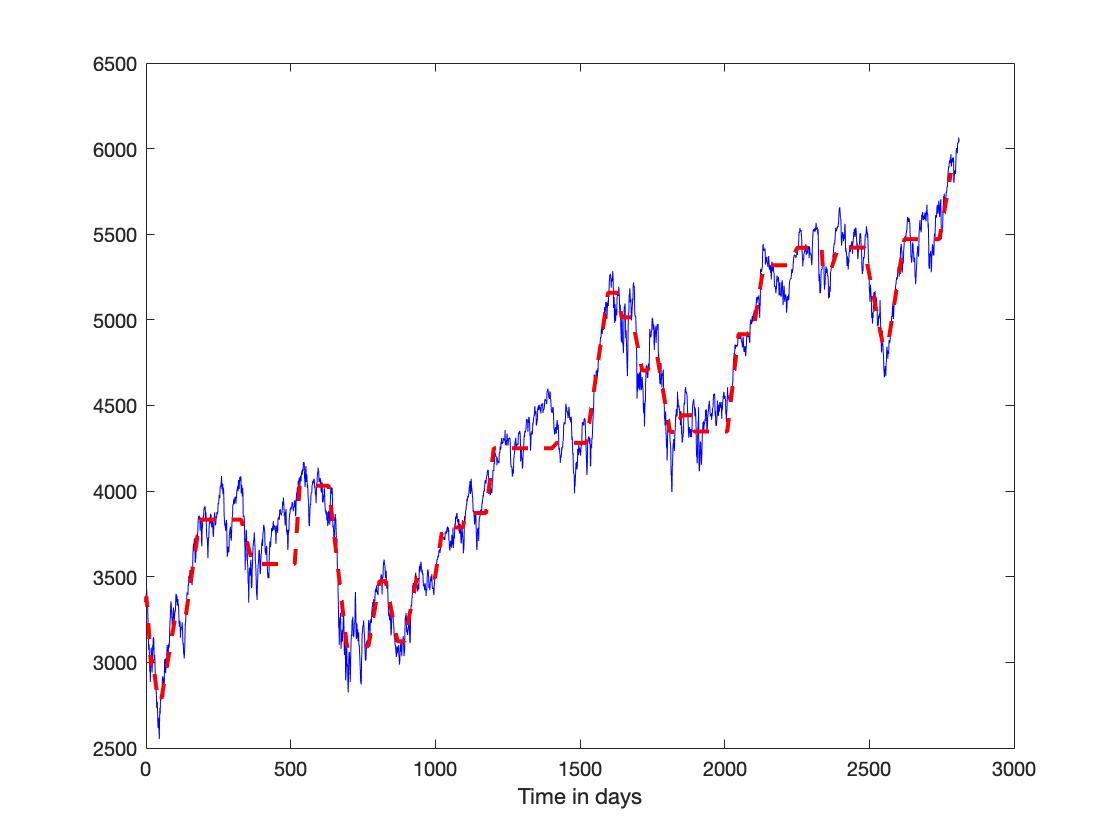,width=.492\textwidth}}
\subfigure[\footnotesize Dérivée estimée (bleu), seuil (rouge - -), passage en $0$ (noir *)]
{\epsfig{figure=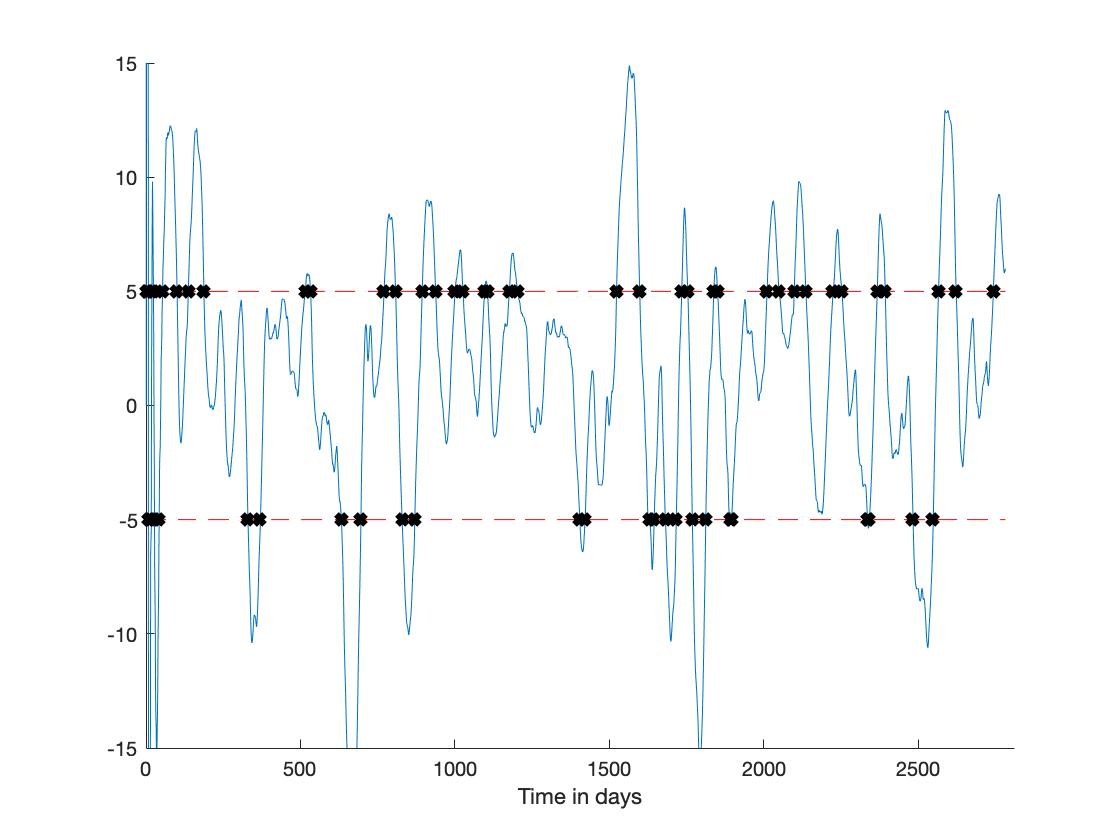,width=.492\textwidth}}
\subfigure[\footnotesize Zoom de (a) ]
{\epsfig{figure=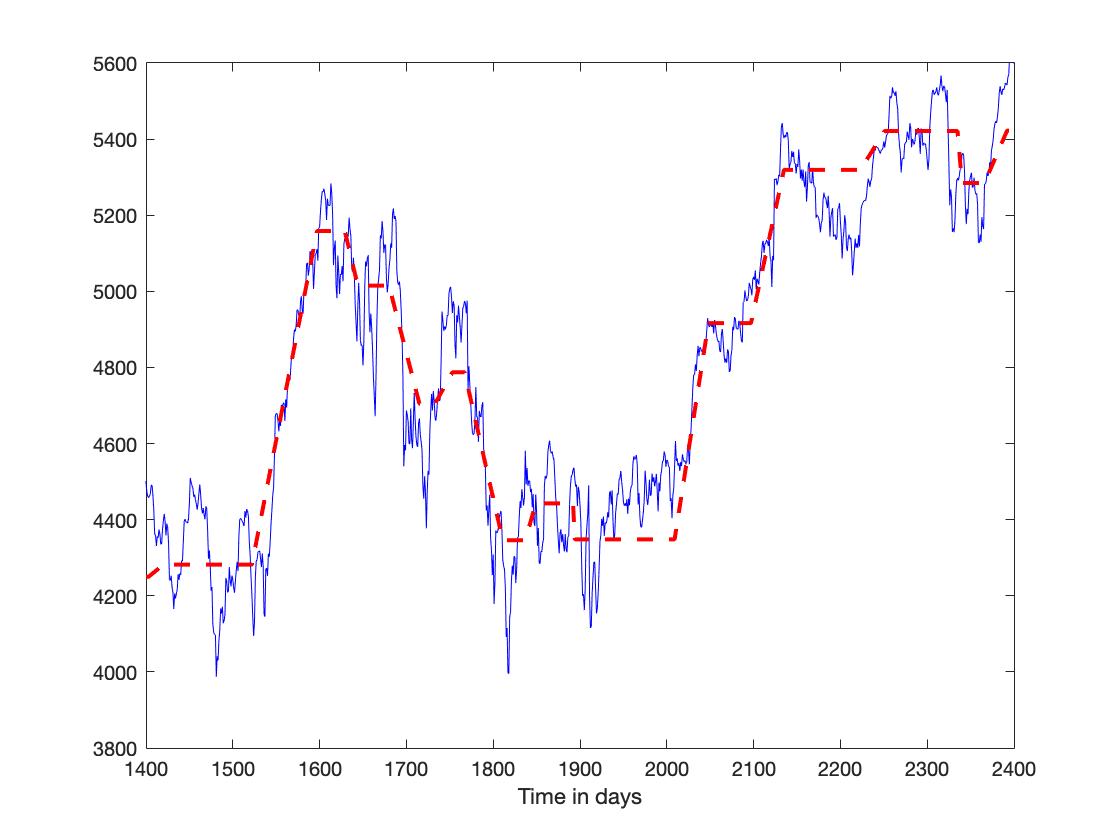,width=.95\textwidth}}
\caption{CAC 40: Seuil haut}\label{S45}
\end{figure*}

On reprend, pour la segmenter, la chronique du CAC 40, représentée par la figure \ref{Si}.  Avec des seuils différents, $1$ et $5$, pour détecter les passages par zéro de la dérivée, on modifie notablement les résultats (figures \ref{S41}-(c) et  \ref{S45}-(c)). 



\section{Conclusion}\label{conclusion}
\begin{enumerate}
\item On examinera d'autres poins d'importance perceptuelle, comme ceux de courbure localement maximale (voir, déjà, \cite{join}).
\item Aux chroniques multivariées correspondent des courbes gauches, c'est-à-dire non planes. Le choix des PIP exige une étude géométrique plus délicate (voir, par exemple, \cite{cagnac}). L'analyse technique, qui examine, souvent, prix et volumes (\cite{bechu,technical,tsinas}), y gagnerait certainement\footnote{Certains liens entre analyse technique, PIP et segmentation des chroniques ont déjà été entrevus (voir, par exemple, \cite{tsinas,wan,wan1,yin}).}.
\item Des applications véritables sont en cours d'étude, la reconnaissance de signatures entre autres. Plusieurs publications proposent, déjà, des techniques diverses d'analyse des chroniques (voir, par exemple, \cite{henniger,hsu}). Nos premiers résultats démontent l'inanité de rechercher mesures ou distances universelles de similarité, c'est-à-dire indépendantes du but poursuivi. 
\end{enumerate}

%
%
%




\newpage

\end{document}